\documentclass[12pt,preprint]{aastex}

\shorttitle{Submillimeter Observations of Ced110}
\shortauthors{Hiramatsu et al.}

\begin{document}

\title{ASTE Submillimeter Observations of a Young Stellar Object Condensation in Cederblad 110}

\author{Masaaki Hiramatsu\altaffilmark{1,2}, Takahiro Hayakawa\altaffilmark{2,3}, Ken'ichi Tatematsu\altaffilmark{2},
 Kazuhisa Kamegai\altaffilmark{4}, Toshikazu Onishi\altaffilmark{5},
Akira Mizuno\altaffilmark{6}, Nobuyuki Yamaguchi\altaffilmark{2}, and Tetsuo Hasegawa\altaffilmark{2}}

\altaffiltext{1}{Department of Astronomy, University of Tokyo, Bunkyo, Tokyo 113-0033, Japan;\\ 
hiramatsu.masaaki@nao.ac.jp}
\altaffiltext{2}{National Astronomical Observatory of Japan, Mitaka, Tokyo 181-8588, Japan}
\altaffiltext{3}{Current address: Department of Physics and Astrophysics, Nagoya University, Chikusa-ku, Nagoya 464-8602, Japan}
\altaffiltext{4}{Institute of Astronomy, University of Tokyo, Mitaka, Tokyo 181-0015, Japan}
\altaffiltext{5}{Department of Physics and Astrophysics, Nagoya University, Chikusa-ku, Nagoya 464-8602, Japan}
\altaffiltext{6}{Solar-Terrestrial Environment Laboratory, Nagoya University, Chikusa-ku, Nagoya 464-8601, Japan}

\begin{abstract}
We present results of submillimeter observations of a low-mass young stellar objects (YSOs) condensation in the
Cederblad 110 region of the Chamaeleon I dark cloud with Atacama Submillimeter Telescope Experiment.
Our HCO$^+$($J$=4-3) map reveals
a dense molecular gas with an extent of $\sim$ 0.1 pc, which is a complex of two envelopes associated
with class I sources Ced110 IRS4 and IRS11 and a very young object Cha-MMS1. The other two 
class I sources in this region, IRS6 and NIR89, are located outside the clump and have no extended 
HCO$^+$ emission. HCO$^+$ abundance is calculated to be $2.6 \times 10^{-10}$ for MMS1 and 
$3.4 \times 10^{-9}$ for IRS4, which are comparable to the reported value for other young 
sources. Bipolar outflows from IRS4 and IRS6 are detected in our $^{12}$CO($J$=3-2) map.
The outflow from IRS4 seems to collide with Cha-MMS1.  
The outflow has enough momentum to affect gas motion in MMS1, although no sign has been detected 
to indicate that a triggered star formation has occurred.
\end{abstract}


\section{INTRODUCTION}
\subsection{Single Star Formation and Cluster Formation}
A number of past observations and theoretical researches have revealed a rough scenario for the formation
of isolated, single low-mass stars.  That is, a gravitational collapse of a molecular cloud core is the 
starting point of star formation, then a protostar with a thick envelope and bipolar outflows
\citep{Bachiller1996} is born. The outflows blow off the protostellar envelope, which could have an impact on 
the mass assembling process of the star \citep[e.g.][]{Arce2002}, affect the 
motion of the cloud around the protostar \citep[e.g.][]{Takakuwa2003}, and cause outflow-triggered star formation 
\citep{Yokogawa2003}, as well as generates turbulence and limit the efficiency of star formation 
\citep{Nakano1995, Matzner2000}. 

It is still veiled how to form stars in a group or a cluster \citep{Lada2003}, although
most stars are known to be born that way.
If stars are forming in a group or cluster, impacts of outflows on the environment 
may be greater than in the formation of isolated stars. Therefore, it is important to study 
interactions between outflows and dense gas in cluster-forming regions for comprehensive understandings
of star formation processes.

\subsection{Cederblad 110 Region}
The Cederblad (Ced) 110 region \citep{Cederblad1946} is located in the center of the Chamaeleon (Cha) I dark cloud, 
one of the low-mass star formation regions in the solar vicinity \citep[$D$=160pc;][]{Whittet1997}. 
\citet{Haikala2005} identified a C$^{18}$O clump in the Ced 110 region, whose radius is 0.19 pc and mass is 11.7 
$M_{\sun }$. This is the most massive clump in their samples. Infrared and millimeter continuum 
observations \citep{Prusti1991, Reipurth1996, Zinnecker1999, Lehtinen2001, Persi2001} have revealed that
seven YSOs at different evolutionary stages and one very young source are clustered within $\sim $0.2 pc. 
In Table~\ref{Ced110table} we list the objects with their position, IR class, $T_{\rm bol}$ \citep{Myers1993},
and $L_{\rm bol}$.

The youngest object in this region, Cha-MMS1, was detected by 1.3 mm continuum observations by 
\citet{Reipurth1996}. At this position, \citet{Lehtinen2001} identified the far-IR object Ced 110 IRS10 by 
observations with the \textit{Infrared Space Observatory} (\textit{ISO}), while no corresponding source has 
been detected with near-IR observations. Thus, MMS1 is considered to be a "real protostar" \citep{Lehtinen2001}. 
However, \citet{Lehtinen2003} could not detect 3 or 6 cm radio continuum emission. This result indicates 
MMS1 does not have any jets or outflows. They noted that MMS1 is still at a prestellar stage.

Ced110 IRS4, a class I object, is associated with a reflection nebula that is extended in a north-south
direction and is divided by
a thick edge-on dust disk \citep{Zinnecker1999, Persi2001}. \citet{Henning1993} detected 1.3 mm radio 
continuum emission from this object, while extended emission was not detected \citep{Reipurth1996}. 

Molecular outflow was detected in this region by $^{12}$CO($J$=1-0) observations with the Swedish-ESO
Submillimeter Telescope \citep[SEST;][]{Mattila1989, Prusti1991}, but the identification of its driving source has been controversial. 
Originally, IRS4 was thought to be driving this outflow. However, \citet{Reipurth1996} detected MMS1 and 
suggested that MMS1 is the driving source of the outflow as well as the nearby Herbig-Haro (HH) objects HH49/50 
\citep[see also][]{Kontinen2000}. While \citet{Lehtinen2003} revealed that MMS1 is associated with no jets,
\citet{Bally2006} suggested that a parsec-scale outflow is ejected from MMS1 based on the positional 
relationship among the HH objects and the YSOs in their large area shock survey. Lack of high-resolution 
mapping observations of the outflow
(e.g. the mapping grid was 1$\arcmin$ in Mattila et al. 1989) have made it difficult to resolve this issue.

Ced110 IRS11 was first detected by \citet{Zinnecker1999} as IRS 4E in their near-IR image.
\citet{Lehtinen2001} tentatively detected a far-IR source from \textit{ISO} data, and identified IRS11 as ISO-ChaI
86 \citep{Persi2000}. The coordinates for IRS11 in Table \ref{Ced110table} are based on the position for
ChaI 86 in \citet{Persi2001}.

In this paper we focus on the following three points; (1) the distribution of dense gas in the YSO clustering 
region, 
(2) the influence of outflows from older protostars on newer ones and the surrounding gas, and (3) the "blue 
asymmetry" line profile toward YSOs which indicates the existence of infalling gas 
\citep{Leung1977, Zhou1992}. 
To investigate them we observed CO($J$=3-2) and HCO$^+$($J$=4-3) lines. The observations of the CO line can 
reveal the distribution of molecular outflows with high spatial resolution and 
enable us to distinguish the driving source of outflows. The HCO$^+$ line has a high 
critical density of $9.7\times $10$^6$ cm$^{-3}$ at the $J$=4-3 transition \citep{Evans1999}, and its observations  
can probe dense gas and also can detect gas infall in protostellar systems through detection of the
blue asymmetry in line profiles. We also made observations of the optically thin H$^{13}$CO$^+$ line 
toward Ced110 IRS4 and Cha-MMS1 to obtain precise physical parameters. In addition, we observed
CH$_3$OH($J$=$7_K$-$6_K$), SiO($J$=8-7), and SO ($J$=$8_9$-$7_8$) lines, whose
upper energy levels\footnote{The data are taken from the Leiden Atomic and Molecular Database (LAMDA).}
are 70.5, 75.0, and 87.5 K, respectively. These high energy levels
made these lines useful diagnostic tools of shocked regions.

We introduce the detail of our observations in \S 2, and the results of the 
observations are described in \S 3. In \S 4 we discuss molecular abundance, gas infall toward protostars, 
kinematics of 
outflows, and the interaction between MMS1 and the outflow from IRS4. Our conclusions are summarized in \S 5.

\section{OBSERVATIONS}
The CO($J$=3-2) and HCO$^+$($J$=4-3) observations were performed in 2004 November with 
the 10 m Atacama Submillimeter Telescope Experiment \citep[ASTE,][]{Ezawa2004}. Observations were remotely
made from the ASTE operation rooms at San Pedro de Atacama, Chile and National Astronomical Observatory of Japan
(NAOJ) Mitaka campus, Japan, using the 
network observation system N-COSMOS3 developed by NAOJ \citep{Kamazaki2005}. We used a 345 GHz (0.8 mm)
band double side band (DSB) superconductor-insulator-superconductor mixer receiver using position switching. 
Taking advantage of the DSB 
observation, we simultaneously received $^{12}$CO($J$=3-2; 345.796 GHz) in the lower side band 
(LSB) and HCO$^+$($J$=4-3; 356.734 GHz) in the upper side band (USB). With the same system, we observed
H$^{13}$CO$^+$($J$=4-3; 346.999 GHz) line in 2005 June and the CH$_3$OH($J$=$7_K$-$6_K$; 338.344 GHz), SO 
($J$=$8_9$-$7_8$; 346.529 GHz), and SiO($J$=8-7; 347.331 GHz) lines in 2006 December.

We used the XF digital spectrometer with a bandwidth and spectral resolution set to 128 MHz and 125 kHz, 
respectively, except for CH$_3$OH line observations made at 512 MHz bandwidth and 500 kHz resolution. The 
velocity resolution
is 0.11 km s$^{-1}$ and 0.43 km s$^{-1}$ in 345 GHz. The half-power beamwidth (HPBW) was 22$\arcsec$ at
345 GHz.

In CO and HCO$^+$ observations, the mapping grid spacing was 20$\arcsec$ (0.016 pc at the distance of
Cha I). According to \citet{Mizuno1994}, the typical size of a star-forming core is about 0.03 pc, 
thus the angular resolution of our
observations is high enough to resolve this size of a core. With this grid spacing, we obtained spectra toward 
136 points which resulted in 4$\arcmin \times 3\arcmin $ maps centered at IRS4. The observations were
made by position switching every 20 or 30 s between the source and off positions. 
The off positions were ($\alpha, \delta$)(J2000) = (11$^{\rm h} 24^{\rm m} 21\fs 2$, 
$-76\arcdeg 57\arcmin 39\farcs 1$) and ($11^{\rm h}42^{\rm m}9\fs7$, $-78\arcdeg 03\arcmin 44\farcs 4$).
 These positions were checked to be free of emission.

We observed H$^{13}$CO$^+$ lines toward Ced110 IRS4 and Cha-MMS1. The CH$_3$OH line was observed
along the line from IRS4 to MMS1 with $20\arcsec$ grid spacing. SO and SiO lines were observed simultaneously
in opposite sidebands
toward the \citet{Kontinen2000} SO northern peak, ($\alpha, \delta$)(J2000) = ($11^{\rm h}06^{\rm m}34\fs7, 
-77\arcdeg 22\arcmin 50\farcs 7$), and the boundary point between IRS4 outflow and MMS1, ($\alpha, \delta$)
(J2000) = ($11^{\rm h}06^{\rm m}34\fs7, -77\arcdeg 23\arcmin 24\farcs 8$). The observed point for the 
shock-tracer molecules are indicated on Fig.\ref{outflowIT}.

DSB system noise temperature at an elevation of 35$\arcdeg $ varied from $T_{\rm sys} =$
300 to 600 K. For CO and HCO$^+$
lines, on-source integration time for each point ranged from 1 to 5 minutes, and the typical rms noise 
was 0.3 and 0.15 K, 
respectively. We made additional integration toward the YSOs for improving signal-to-noise ratio in the 
HCO$^+$ line, reaching a rms noise of 0.1 K. For other lines, we made the rms noise level go down to 0.05 K for 
H$^{13}$CO$^+$, 0.1 K for CH$_3$OH, 0.15 K for SiO, and 0.2 K for SO. Despite a careful inspection,
no emissions from shock-tracing molecules such as CH$_3$OH, SiO, and SO were detected in our observations.
All temperatures are reported in 
terms of the $T_{\rm MB}$ scale. The main beam efficiency was $\eta_{\rm MB} \sim 0.6$.

The telescope pointing was regularly checked by observing Jupiter, Saturn, or IRC+10216 every 1 or 2 hr during 
observations, and the error was estimated to be about 5$\arcsec $ peak to peak. Intensity calibration
was carried out by the chopper-wheel method \citep{Kutner1981}. The absolute intensity is calibrated
by assuming that $T_{\rm MB}$ of IRC+10216 for CO ($J$=3-2) is 32.5 K \citep{Wang1994}. The daily
variability was checked by observing L1551 IRS5 for HCO$^+$ and by observing Ori KL for H$^{13}$CO$^+$, 
CH$_3$OH, SO, and SiO.
The sideband ratio was assumed to be unity. The zenith 220 GHz opacity varied from $\tau \sim
0.05$ to $\sim 0.1$. \citet{Matsuo1998} reported the atmospheric transmission is 0.86 and 0.83 in 345 and
356 GHz, respectively, when the zenith 220 GHz opacity was 0.04. In this paper, we assume the atmospheric 
transmissions in the two side-bands are identical.

We reduced the data using the NEWSTAR software developed in the Nobeyama Radio Observatory. The baseline
subtraction was done by a straight line in most cases.

\section{RESULTS}
\subsection{Overview}
Figures \ref{outflowIT} and \ref{outflowCM} show the observed maps of the CO($J$=3-2) emission. The CO($J$=3-2)
emission was detected at all points within the 4$\arcmin \times 3 \arcmin $ map
with a varying degree of intensity. Around IRS4 and IRS6, we detected CO high-velocity components 
most probably caused by outflows. The details of the outflows are described in \S 3.3 and \S 4.3. 

Figure \ref{HCOmap} shows the integrated intensity map of the HCO$^+$($J$=4-3) line in the velocity 
range of 2.0 km s$^{-1}$ $\le  V_{\rm LSR} \le $ 7.0 km s$^{-1}$. A HCO$^+$ clump of dense gas is seen
in Fig.\ref{HCOmap}. This clump is an aggregate of three independent envelopes associated with Ced110
IRS4, IRS11, and Cha-MMS1. The envelope associated with IRS11 seems to be apart from 
the component associated with IRS4 in the $V_{\rm LSR} =$ 4.78 and 5.00 km s$^{-1}$ panels of Fig.\ref{HCOchmap}.
MMS1 is also distinguishable in the $V_{\rm LSR} =$ 3.92 and 4.13 km s$^{-1}$ panels. Figure \ref{pv7} and 
\ref{pv9} are the position-velocity diagrams of CO and HCO$^+$ lines along line A and B in Fig.\ref{HCOmap}, 
respectively. In Fig.\ref{pv7}
the components associated with IRS4 and IRS11 are marginally resolved with the viewpoint of line width, that is, 
the emission associated with IRS4 exhibits a distinctly larger linewidth than that for IRS11. 
The HCO$^+$ clump covers $140\arcsec \times 110\arcsec$ (0.11 pc $\times$ 0.085 pc) above the 3 $\sigma$ noise 
level. The extension of the clump with a position angle $\sim 45$ deg corresponds well to the 200 $\mu$m ridge
emission observed with \textit{ISO} \citep{Lehtinen2001}. 

\subsection{CO and HCO$^+$ Line Profiles}
\begin{figure}
\epsscale{.80}
\plotone{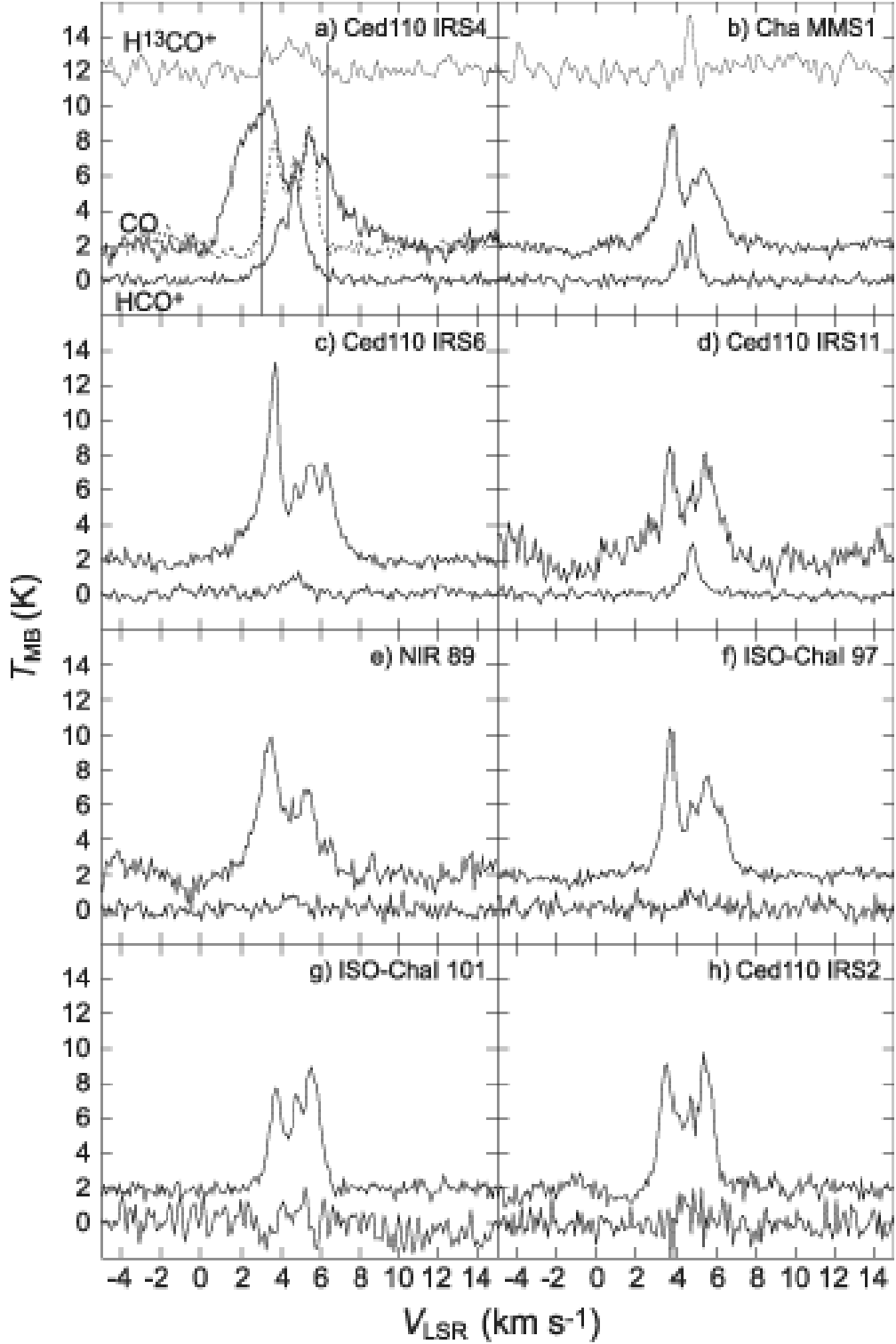}
\caption{HCO$^+$($J$=4-3) (multiplied by a factor of 2), CO($J$=3-2) (with 2 K offset), 
and H$^{13}$CO$^+$($J$=4-3) (only in IRS4 and MMS1 panels with 12 K offset and multiplied by a 
factor of 15) line profiles 
toward Cha-MMS1 and YSOs in Ced110 region. In panel \textit{a}, the spectrum toward the edge of the map is 
indicated with a dotted line, and the borders between outflow and line core are shown with two vertical 
lines. We identified $V_{\rm LSR} \le$ 3.0 km s$^{-1}$ and $V_{\rm LSR} \ge$ 6.3 km s$^{-1}$ as outflow. 
\label{YSOlines}}
\end{figure}

The CO($J$=3-2), HCO$^+$($J$=4-3), and H$^{13}$CO$^+$($J$=4-3) line profiles obtained
at YSO positions are shown in Fig.\ref{YSOlines}. The CO line profile from ambient gas is sampled at the 
boundary of the map and shown in Fig.\ref{YSOlines}\textit{a} by a dotted line.
 
Almost all CO line profiles have a dip at $V_{\rm LSR} \sim 4.3$ km s$^{-1}$. The dip also appears in 
some HCO$^+$ line profiles at the same velocity. The velocity of the dip is identical to 
that of C$^{18}$O clump number 3 identified in the Ced110 region by \citet{Haikala2005}, and also gives close
agreement with the velocities of optically thin lines observed by \citet{Kontinen2000}. These correspondences
mean that this dip is caused by self-absorption by less excited gas, not the result of two emission 
components with different velocities along the same line of sight. Since the observed region corresponds to one 
C$^{18}$O clump, the outer gas of the clump in the lower excitation level would make the CO absorption.
The dip in the HCO$^+$($J$=4-3) line indicates a significant population of the $J=3$ level that requires a
moderately high density, but with a lower $J$=4-3 excitation temperature than in the center of the clump. This
is a reflection of the density gradient in the clump. We can find these dips in Fig.
\ref{pv7} and \ref{pv9}. This dip prevents the channel map 
(Fig.\ref{HCOchmap}) from representing the actual distribution of interstellar matter in the dip velocity.
The indentations of contours located at the south of IRS4 and a lack of a peak around MMS1 in the 
panels of $V_{\rm LSR}= 4.35$ and 4.57 km s$^{-1}$ of Fig.\ref{HCOchmap} may be due to the self-absorption. 
It is conceivable that other reasons such as depletion and chemistry in the dense gas cause this lack of a
peak. The issue of molecular depletion is discussed in \S 4.1.

The parameters of the lines detected at YSO positions are summarized in Tables \ref{COtable}, \ref{HCOtable} 
and \ref{H13COtable}. We fitted the spectral data of HCO$^+$ and H$^{13}$CO$^+$ with a Gaussian profile. 
The error estimates of the values in the tables
come from the fit. In deriving the line parameters, the velocity ranges of 
outflows and self-absorption features in the lines are eliminated in the fitting processes. Although this may
have some defectiveness, this is the optimal handling of our data.

\subsection{Individual Sources}
\subsubsection{Ced110 IRS4 and the Outflow}
Bipolar outflow from IRS4 is detected in our CO($J$=3-2) observations. This outflow had been also 
detected by \citet{Mattila1989} and \citet{Prusti1991} with their CO($J$=1-0) observations. 
We compared the line profiles toward IRS4 with that of the ambient gas obtained at the edge of the 
map in ($\alpha, \delta$)(J2000) = ($11^{\rm h}06^{\rm m}24\fs3$, $-77\arcdeg 21\arcmin 16\farcs 7$). 
Judging from this comparison shown in 
Fig.\ref{YSOlines}\textit{a}, the line profile
in the velocity ranges of $V_{\rm LSR} \le 3.0$ km s$^{-1}$ and $V_{\rm LSR} \ge $6.3 km s$^{-1}$ are not 
affected by the ambient gas, and thus we identified these ranges as outflow components.

\begin{figure}
\epsscale{1.0}
\plotone{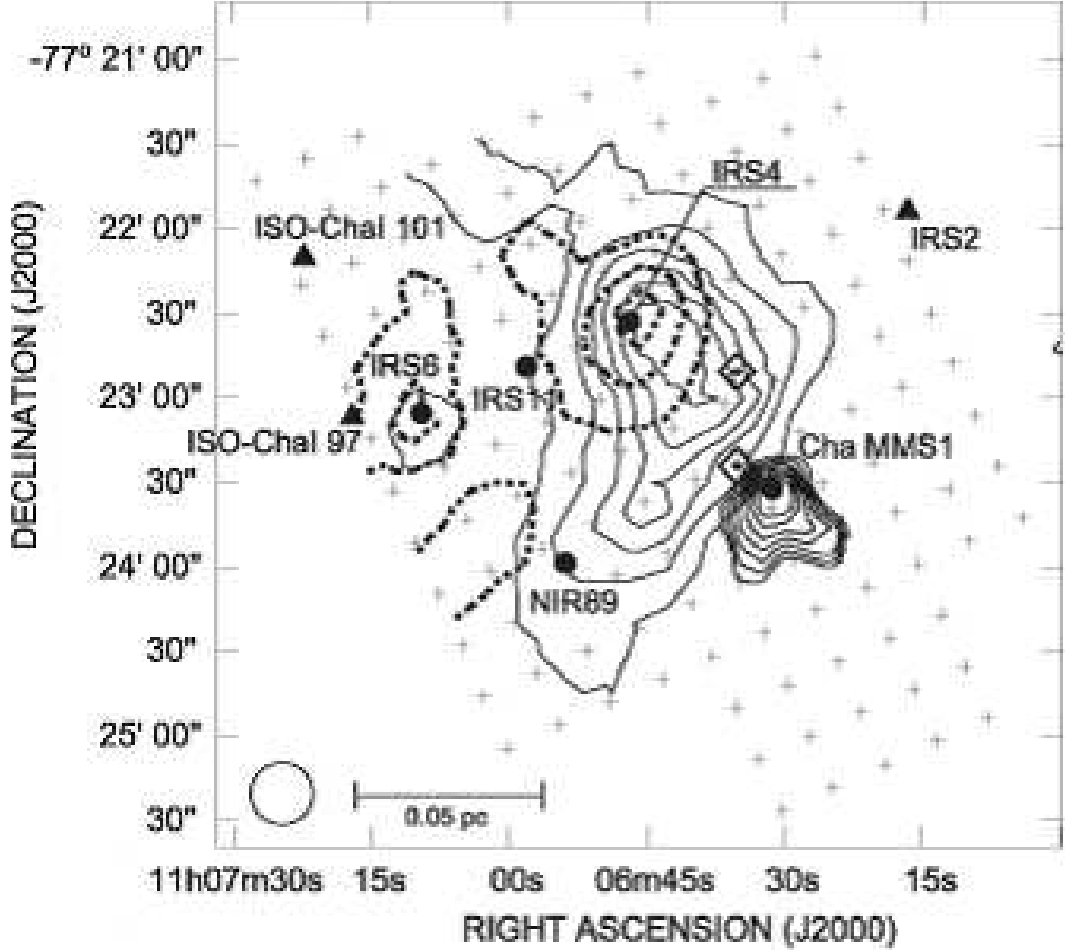}
\caption{Distribution of high velocity component of CO($J$=3-2) line. The blueshifted (-4.0 km s$^{-1}$
$\le V_{\rm LSR} \le 3.0$ km s$^{-1}$) and the redshifted (6.3 km s$^{-1}$ $\le V_{\rm LSR} \le$ 10.3 km s$^{-1}$)
components are drawn with solid and dotted lines, respectively.
 The first contour level and contour interval are 1.53 K km s$^{-1}$ for the blueshifted components and 
1.1 K km s$^{-1}$ for redshifted components. Cha-MMS1 observed with a 1.3 mm continuum 
\citep{Reipurth1996} is also shown with thick lines. The first contour level is 200 mJy, and the interval
is 25 mJy.  Small
crosses indicate observed points. Each YSO is shown by filled circles (pre-stellar core and class I) or 
triangles (class II or III). The diamonds with the central dot denote the observed points of SO and SiO lines.
The HPBW is indicated with an open circle.
\label{outflowIT}}
\end{figure}

\begin{figure}
\epsscale{.80}
\plotone{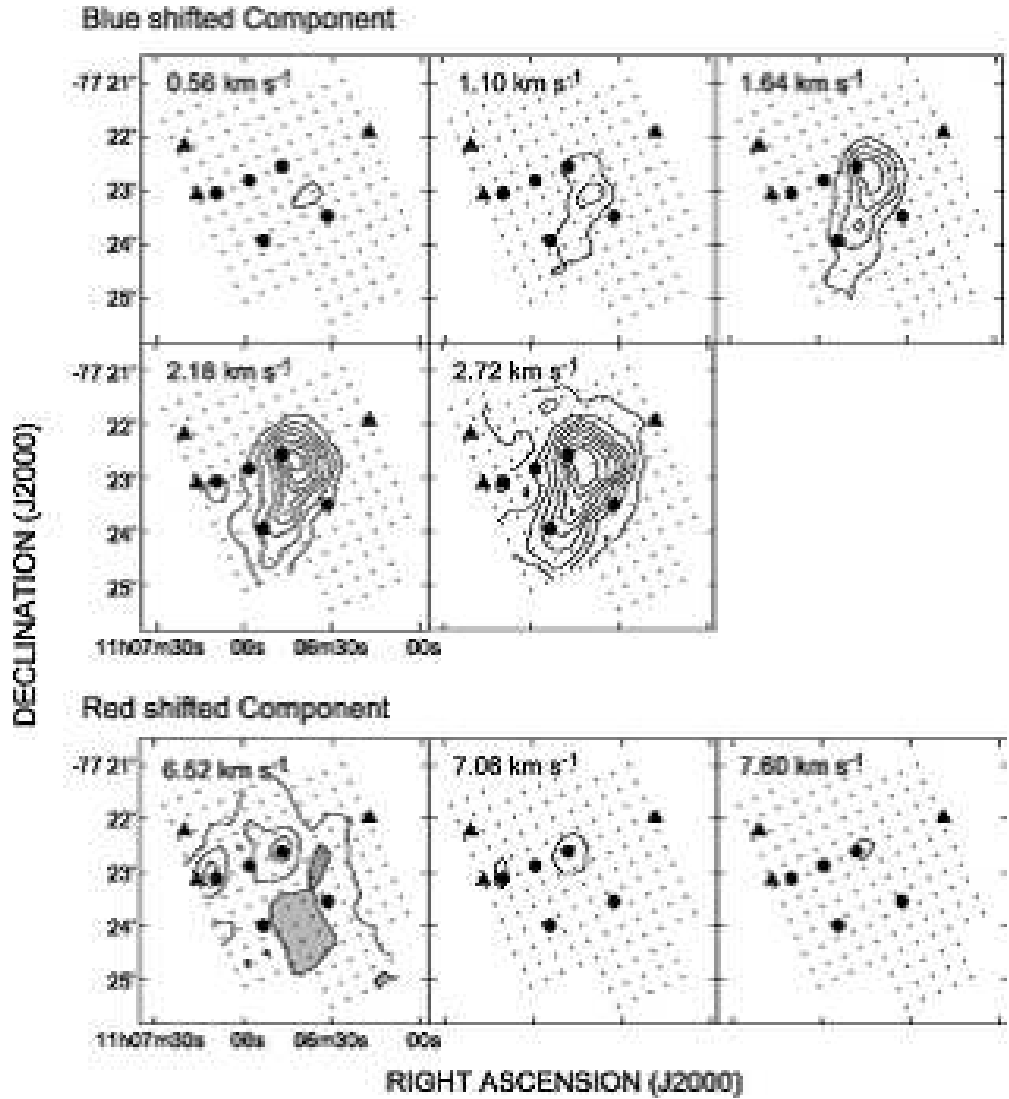}
\caption{Velocity channel maps of CO($J$=3-2) in the outflow range. The central LSR velocity for each
integral in units of km s$^{-1}$ is shown at the top left corner of each panel. The first contour level and 
contour interval are 0.8 K. The markers are the same as in Fig.\ref{outflowIT}. The gray region represents local
minimum.
\label{outflowCM}}
\end{figure}

\begin{figure}
\epsscale{.80}
\plotone{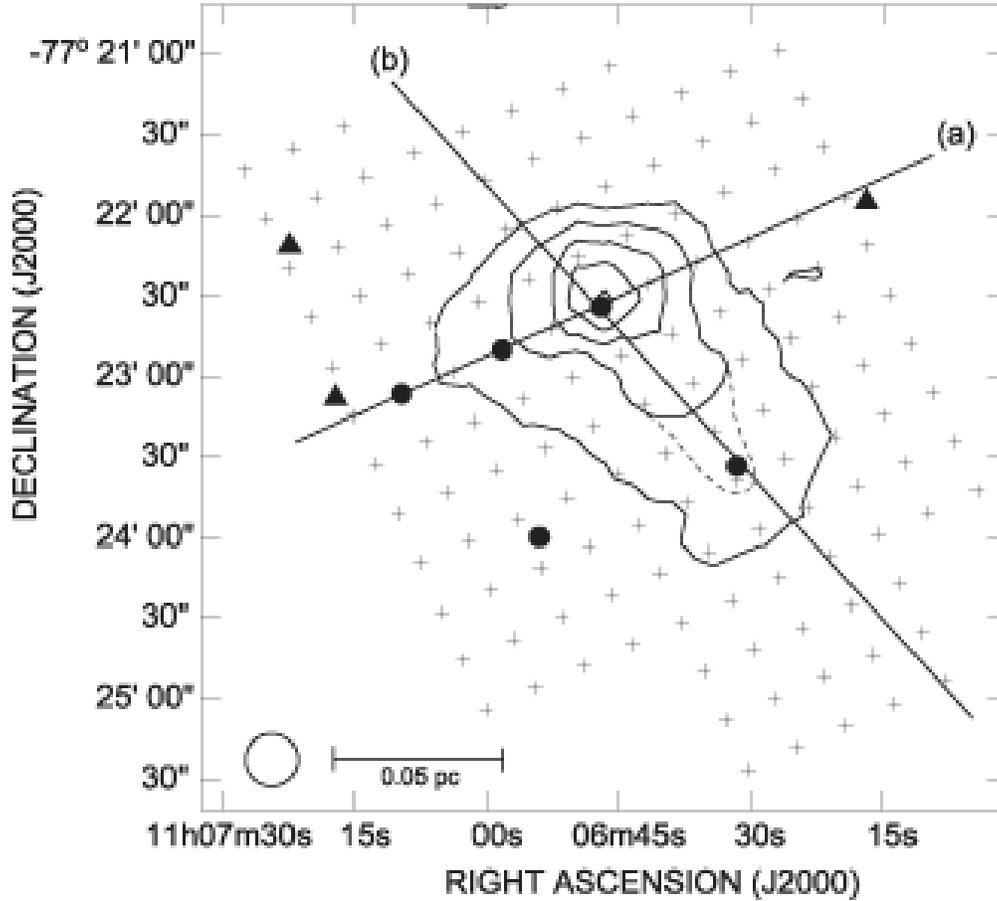}
\caption{Integrated intensity map of Ced110 in HCO$^+$($J$=4-3) and the distribution of YSOs.
The integrated velocity range is 2.0 km s$^{-1}$ $\leq V_{\rm LSR} \leq$ 7.0 km s$^{-1}$. The first contour level and
contour interval are 0.8 K km s$^{-1}$. Dashed line shows the "absorption-free" integrated intensity
contour (See \S 3.3.2). The open circle shows the HPBW. Filled circles and triangles indicate the 
positions of YSOs as in Fig.\ref{outflowIT}. The straight lines are also shown along which the position-velocity 
diagrams in Fig.\ref{pv7} and \ref{pv9} are taken.
\label{HCOmap}}
\end{figure}

\begin{figure}
\epsscale{.80}
\plotone{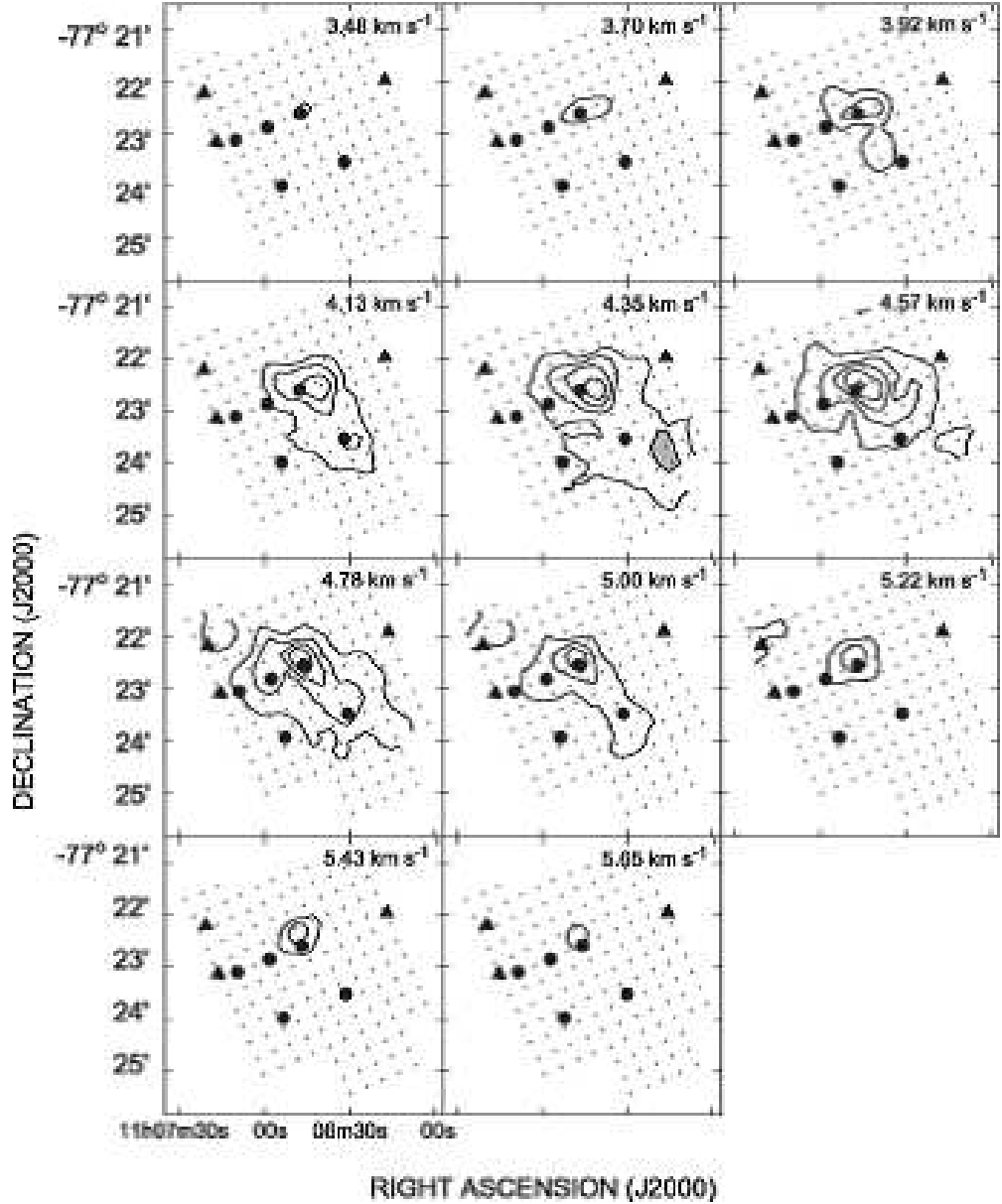}
\caption{Velocity channel maps of HCO$^+$($J$=4-3) transition. The central LSR velocity for each
integral in units of km s$^{-1}$ is shown at the top right corner of each panel.
The first contour level and contour interval are 0.5 K. The positions of YSOs
are indicated with the same symbols as in Fig.\ref{outflowIT}. The gray region represents local minimum.
\label{HCOchmap}}
\end{figure}

\begin{figure}
\epsscale{.80}
\plotone{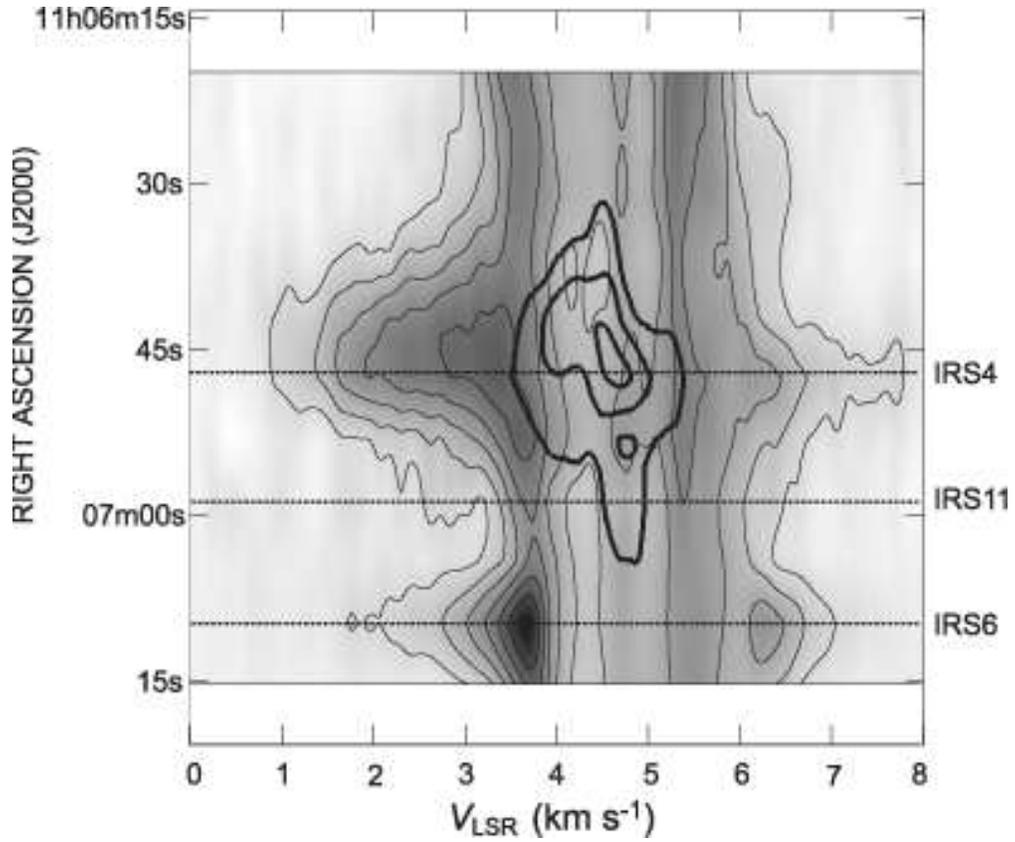}
\caption{Position velocity diagram along the line (a) in Fig. \ref{HCOmap}. CO($J$=3-2)
and HCO$^+$($J$=4-3) are shown with thin and thick line, respectively. The first 
contour level and contour interval are 1.5 K for CO and 0.9 K for HCO$^+$. The positions of IRS4, IRS6, and IRS11
are indicated with dashed lines.
 \label{pv7}}
\end{figure}

The distribution of the outflow components from IRS4 are shown in Fig.\ref{outflowIT} (integrated 
intensity map) and Fig.\ref{outflowCM} (channel map). The blueshifted and
redshifted outflows show a different aspect. The spatial extent of the blueshifted outflow is considerably 
larger than that of the 
redshifted outflow. Furthermore, the highest velocity component of the blueshifted outflow lies 
midway between IRS4 and MMS1, while the velocity reaches its peak around IRS4 in the redshifted flow.
The near-IR images \citep{Zinnecker1999, Persi2001} indicate that IRS4 is seen nearly 
edge-on. The observed distribution of the blueshifted component of the outflow extending to the south 
suggests that the outflow axis slightly inclines with the southern side closer 
than the northern side. This configuration is in agreement with the previous remarks on the inclination of IRS4
based on the color gradient of the nebulae \citep{Zinnecker1999}. They found that the northern part of the 
reflection nebula is redder than the southern part, which suggests that the northern part is tilted away from us.
Our integrated intensity map (Fig.\ref{outflowIT}) does not show the redshifted component in the south of MMS1
reported in the literature \citep{Mattila1989, Prusti1991}. In the channel map (Fig.\ref{outflowCM}), however, 
one can see the extended redshifted emission at $V_{\rm LSR}=6.52$ km s$^{-1}$, and this could be its 
counterpart. 

Observations of the H$^{13}$CO$^+$($J$=4-3) line were also performed toward IRS4 (Fig.\ref{YSOlines}a)
with a 43 mK noise level. The FWHM of this line, $\Delta V = 1.72 \pm
0.29$ km s$^{-1}$, corresponds to that of the HCO$^+$ line within the margin of error. This supports 
the detection of the line. With this result, we discuss molecular abundance in \S 4.1.

\subsubsection{Cha-MMS1}
In Fig.\ref{YSOlines}\textit{b} we show the CO, HCO$^+$, and H$^{13}$CO$^+$ line profiles toward Cha-MMS1. 
The dip in the HCO$^+$ line is deeper than that of IRS4. This deeper dip shows that MMS1 is deeply embedded in
the densest part of C$^{18}$O clump number 3 \citep{Haikala2005}, because MMS1 lies nearer to the position of 
the peak intensity of C$^{18}$O than IRS4.

We calculated the HCO$^+$ "original" integrated intensity with a Gaussian
fit using the line profiles in the velocity ranges which are free from self-absorption. These reconstructed 
"absorption-free" contours are shown with dashed linea in Fig.\ref{HCOmap}. The summit ridge of the contours extends
south and reaches the position of MMS1. This result again supports the idea that MMS1 is deeply
embedded and its outer envelope makes this absorption feature.
  
Around MMS1, there are
no conspicuous wide lines (Fig.\ref{pv9}). This shows that no outflow is associated with MMS1, which
is consistent with the lack of centimeter continuum emission from ionized jets \citep{Lehtinen2003}. 
Thus, MMS1 is rejected as a driving source of the outflows seen in CO ($J=1-0$) maps \citep{Mattila1989,
Prusti1991} as well as a launcher of HH49/50.

There are some literatures which summarized properties of class 0 objects 
\citep{Greg1997, Mardones1997, Greg2000, Froebrich2005}. We 
performed a literature survey about three aspects of the class 0 objects: centimeter continuum emission, CO
outflows, and HH objects. All class 0 sources that appeared in the four papers had at least one aspect out of
the three. On the other hand, MMS1 has no outflow, centimeter continuum emission, or HH objects. 
Our results show no evidence of protostar activity in MMS1.

The H$^{13}$CO$^+$($J$=4-3) line was detected toward MMS1. The width of the line, 0.36 km s$^{-1}$, is 
smaller than that of the same line toward IRS4. This optically thin line could reflect the cold innermost 
region of MMS1; thus, the narrow line shows there are no protostars formed yet in MMS1.

The virial mass of MMS1 is derived from the H$^{13}$CO$^+$ line width. Assuming the diameter of MMS1
is $40'' = 0.03$ pc \citep{Reipurth1996}, the virial mass is calculated to be 0.49 $M_{\sun }$. This is 
comparable to the mass (0.45 $M_{\sun }$) derived by \citet{Lehtinen2001} from 1.3 mm continuum observation.

\subsubsection{Ced110 IRS6}
IRS6 is a binary source (IRS6a and IRS6b), and the near-IR luminosity of IRS6a is a factor of 7 higher than the
companion \citep{Persi2001}. This fact indicates that the features obtained toward the binary system mainly 
derive from IRS6a. 

Toward IRS6, we found a weak outflow component in the CO line (Fig.\ref{outflowIT}, \ref{pv7}, 
and \ref{YSOlines}\textit{c}) which has not been detected in past radio observations. The 
redshifted component extends north, while the blueshifted component is tightly associated with IRS6. The 
kinematics of the outflow is described in \S 4.2.

We marginally detected a weak HCO$^+$ line toward IRS6 (Fig.\ref{YSOlines}\textit{c}). In Fig.\ref{HCOchmap} 
and \ref{pv7}, the IRS6 system does not seem to be surrounded by an extended HCO$^+$ envelope. 
This result supports previous study by \citet{Henning1993} that IRS6 has no massive circumstellar envelope.
The small envelope and high $T_{\rm bol}$ of 270 K suggest IRS6 is a relatively evolved protostar.

\subsubsection{Ced110 IRS11}
Toward IRS11, no high-velocity components were detected by either CO($J$=3-2) or HCO$^+$($J$=4-3) lines.
A small dense gas envelope associated with this object is detected with the
HCO$^+$ line (Fig.\ref{HCOmap}, \ref{HCOchmap}, and \ref{pv7}). The 
radius of the envelope appearing in Fig.\ref{pv7} is about 20$''$, which corresponds to 1.6 $\times 
10^{-2}$ pc at the distance of the Cha I dark cloud. 

\subsubsection{NIR89 and Class II/III Sources}
Though NIR89 is classified as a potential class I \citep{Persi2001}, no CO($J$=3-2) outflow was
detected (Fig.\ref{YSOlines}\textit{e}). In addition, no HCO$^+$($J$=4-3) emission was detected
toward the source. 
Because this object is thought of as a "good candidate young brown dwarf" with a stellar mass of 0.02 
$M_{\sun }$ \citep{Persi2001}, the amount of the envelope associated with NIR89 should be too small to be
detected by our HCO$^+$ observations.

No CO outflow components or HCO$^+$ emission were detected toward the class II/III objects, ISO-Cha I
97, ISO-ChaI101, and Ced110 IRS2 (Fig.\ref{YSOlines}\textit{f - h}). This lack of outflow is common among T Tauri
stars in general. For HCO$^+$ emission, \citet{Zadelhoff2001} 
detected the line toward classical T Tauri star LkCa 15 in the Taurus region with a peak $T_{\rm mb}$ of 0.14 K
using the James Clerk Maxwell Telescope. Although the line intensity could vary depending on the inclination 
of the circumstellar disks, in the case of class II/III objects, it is still plausible that
we could not detect the HCO$^+$ line toward those samples in our observations with 0.15 K sensitivity 
($1 \sigma$).

\section{DISCUSSION}
\subsection{Molecular Abundance in Envelopes}
In dense, cold gas envelopes of protostars and prestellar cores, molecules often freeze out onto dust grains
\citep[see e.g.][]{Jorgensen2005}. On the other hand, the abundance of some species like HCO$^+$ are known 
to be enhanced in outflows \citep{Rawlings2004}. 

We examined the abundance of HCO$^+$ toward MMS1 and IRS4 by means of our H$^{13}$CO$^+$ ($J=4-3$)
observations and
1.3 mm continuum observations \citep{Henning1993, Reipurth1996}. Assuming that the dust emission
is optically thin, the column density of the H$_2$ molecule, $N$(H$_2$), is calculated with
\begin{eqnarray}
	N({\rm H}_2) = {{S^{\rm beam}_{\nu}} \over \Omega _{\rm beam}\mu m_{\rm H} \kappa_\nu B_\nu
	(T_{\rm D})},
	 \label{dust2N}
\end{eqnarray}
where $S^{\rm beam}_{\nu}$ is the 1.3 mm flux density per beam, $\Omega _{\rm beam}$ is the beam solid angle, 
$\mu=2.3$ is the mean molecular weight, $m_{\rm H}$ is the mass of the H atom, $\kappa_{\rm 1.3 mm}$
is the mass absorption coefficient per gram, and $B_\nu (T_{\rm D})$ is the Planck function at the
dust temperature $T_{\rm D}$. We employed $\kappa_{\rm 1.3 mm}=0.01$ cm$^2$ g$^{-1}$ for IRS4, which is the
recommended value for very dense regions \citep{Ossenkopf1994}. Meanwhile, we applied the value of 0.005 cm$^2$
g$^{-1}$ for MMS1, which is the adopted value for prestellar core L1689B \citep{Andre1996}. The substituted
values of $S^{\rm beam}_{\nu}$ are 370 mJy for MMS1 \citep{Reipurth1996} and 101 mJy for IRS4 
\citep{Henning1993}. We employed a dust temperature of 20 and 21 K for MMS1 and IRS4, 
respectively, based on \citet{Lehtinen2001}. The resulting H$_2$ column density is $2.2 \times 10^{23}$
cm$^{-2}$ for MMS1 and $2.8 \times 10^{22}$ cm$^{-2}$ for IRS4. 

We also calculated the column density of H$^{13}$CO$^+$ under LTE conditions and the assumption that the
excitation temperature equals the dust temperature. Comparing $N$(H$_2$) obtained from dust observations 
with $N$(H$^{13}$CO$^+$), we estimated the abundance of HCO$^+$ to be $X$(HCO$^+$) = $2.6 \times 10^{-10}$ for 
MMS1 and $3.4 \times 10^{-9}$ for IRS4. We set the isotropy ratio of [H$^{13}$CO$^+$]/[HCO$^+$] = 70. 
The results are summarized in Table \ref{NH2table}.

\citet{Jorgensen2004} reported average $X$(HCO$^+$) $= 8.0 \times 10^{-10}$ for prestellar cores and 
$1.1 \times 10^{-8}$ for class I sources. \citet{Jorgensen2005} pointed out that the molecular
depletion in envelopes of class I objects is less significant because of heating by their central sources.
Furthermore, \citet{Rawlings2004} reported that the abundance of HCO$^+$ is enhanced in molecular outflows. 
Our results are not in contradiction with abundances reported in similar objects. However, it is not
appropriate to investigate the results in more detail, because the mass absorption coefficient $\kappa$ 
and the LTE approximation have large uncertainties. $\kappa$ has different values depending on dust size,
shape, structure, and so on \citep{Henning1995}. In addition, \citet{Mennella1998} show $\kappa$ has a 
dependence on temperature.

\subsection{Gas Infall toward the Protostars}
We show HCO$^+$($J$=4-3) line profiles toward four class I sources in Fig. \ref{YSOlines}\textit{a}, 
\textit{c}, \textit{d} and \textit{e}. 
None of the profiles show the blue asymmetry which is one of the indicators of infalling gas toward protostars
\citep{Leung1977, Zhou1992}. In a simplified model of an infalling envelope, the blueshifted line
component is stronger than the redshifted component. This is because the redshifted component
is absorbed by the cooler infalling material in the near half of the envelope along the line of sight.
Do our results mean these objects have no infalling gas? 

As mentioned in the previous sections, IRS4 and IRS6 obviously have outflows. This is clear
evidence of the existence of infalling gas toward the protostars. \citet{Greg1997,Greg2000} have revealed that
blue asymmetry line profiles do not appear in about half of class 0 and I sources, although they are thought 
to be in the active accretion phase.

Some reasons are suggested why blue asymmetry does not appear in all sources. One great problem is 
confusion with outflows. If a redshifted outflow exists in the telescope beam, the redshifted line components 
absorbed by the infalling envelope would be compensated. It is also difficult to separate the influence of other 
velocity components such as circumstellar disks and ambient gas in observations with a large beam of single 
dish radio telescopes. 
In our observations, as an example, a wing component is detected in the CO line profile toward IRS4. This 
profile obviously shows that our 22$''$ beam covers the outflowing component.
In general, there are some possibilities to confuse 
not only outflow, but core rotation and other large-scale physical structures with infall, all of which affect
obtained line profiles. Thus, it is difficult to investigate infall motion in targeted observations toward 
protostars without high-resolution spatial information about the envelope. 

The other problem is optical depth. Choosing molecular lines with appropriate optical depth is necessary to 
detect infall motion of the inner portion of the envelope in general, not only for Ced110. When a protostar is 
highly embedded in a dense clump and an observed line becomes optically too thick, the line cannot convey
information of the inner infalling region any more and reflects only properties of foreground components.

\subsection{Kinematics of Outflows}
We calculated the physical parameters of the outflows from IRS4 according to 
the procedures of \citet{Choi1993}. In these procedures, the CO column density for each channel in units of 
cm$^{-2}$, $N^{\rm CO}_i$, and the mass per channel, $M_i$, are calculated from 
\begin{eqnarray}
	N^{\rm CO}_i &=& 1.10 \times 10^{15}{T_{\rm R 3-2}\Delta V \over D(n, T_{\rm K})}{\tau _{32} \over 
	1-\exp(-\tau _{32}) },\\
	M_i &=& \mu m_{{\rm H}_2} d^2 \Omega N^{\rm CO}_i {[{\rm H}] \over [{\rm C}]}{[{\rm C}] \over [{\rm CO}]},
	 \label{outflow}
\end{eqnarray}
where
\begin{eqnarray}
	 D(n, T_{\rm K}) = f_2[J_\nu (T_{\rm ex})-J_\nu (T_{\rm bk})][1-\exp(-16.597/T_{\rm ex})],
	 \label{}
\end{eqnarray}
$T_{\rm R 3-2}$ is the line intensity of the CO($J$=3-2) line, $\Delta V$ is the channel width in km s$^{-1}$,
$f_2$ is the fraction of CO molecules in the $J=2$ state, $d$ =
160 pc is the distance to the Cha I cloud, $\Omega$ is the solid angle subtended by emission, [H]/[C]$= 2.5 
\times 10^3$, [C]/[CO] $= 8$, and $\tau _{32}$ is the optical depth of the line. We calculated $\tau _{32}
\sim 0.4$ comparing with CO($J$=3-2) and CO($J$=1-0) line-wing intensity \citep{Prusti1991}. In this calculation,
we assumed 
that the excitation temperatures $T_{\rm ex}$ of the two lines are equal to 20 K. According to large velocity
gradiant simulations by \citet{Choi1993}, $D(n, T_{\rm K})$ does not
vary so much within the reasonable density and temperature range. Therefore, we set the same value of 
$D(n, T_{\rm K}) = 1.5$ as in \citet {Choi1993}. The mass $M$, momentum $P$, and kinetic energy $E_{\rm k}$ 
can be estimated from
\begin{eqnarray}
	M &=& \Sigma _i M_i, \\
	P &=& \Sigma _i M_i |V_i - V_0|, \\
	E_{\rm K} &=& \Sigma _i {1 \over 2} M_i (V_i - V_0)^2, 
\end{eqnarray}
where $V_0$ is the system velocity. We also calculated the dynamical timescale $t_{\rm d}= R / V_{\rm ch}$
where the characteristic velocity $V_{\rm ch} = P / M$, mass loss rate $\dot{M} = M/t_{\rm d}$, and average 
driving force $F = P / t_{\rm d}$.

The results are shown in Table \ref{outflowtables}. We adopted an inclination of the outflow axis from the 
line of sight of $i =72 \arcdeg$ \citep{Pontoppidan2005}. They derived this value through comparing 
near-IR images with three-dimensional Monte Carlo radiative transfer code of a disk-shadow projection model.
This value is consistent with the fact that the high-resolution infrared images 
\citep{Zinnecker1999, Persi2001} clearly show the edge-on disklike silhouette of IRS4. 

We compared the derived values with those in a statistical study of outflows \citep{Wu2004} in which 
they show
a correlation between the mass, $L_{\rm bol}$ and $F$ for 391 samples. The mass of IRS4's blueshifted
outflow is smaller by about 1 order of magnitude than the average of the same $L_{\rm bol}$ sources
in \citet{Wu2004}, but it is still within the range of variation of the data.

In contrast, the derived mass loss rate is rather large in IRS4 despite its small size. Generally, 
the mass loss rate by jets is about 1\% and that by outflows is10 \% of the mass accretion rate 
\citep{Hartigan1995}. The derived value of 1.1 $\times 10^{-6} M_{\sun }$ yr$^{-1}$ is comparable to 
the typical accretion rate for protostars, 10$^{-6} M_{\sun }$ yr$^{-1}$, derived by the similarity
solution for thermally supported clouds with 10 K \citep{Hartmann1998}. 

HH49/50 is a group of knots located 10.5$\arcmin$ southwest of the Ced110 region. Studies of radial velocity 
and proper motion \citep{Schwartz1980, Schwartz1984} revealed the knots move toward the southwest and away 
from us. \citet{Mattila1989} and \citet{Prusti1991} identified Ced110 IRS4 as the driving source of 
these HH objects judging from their CO outflow map, because they found a diffuse redshifted component
that extends
in the southwest of IRS4 and the alignment of knots is almost parallel to the direction to IRS4.
In our observations, we detected two components that extend to the south of IRS4. One is the prominent blueshifted
outflow which has an opposite line-of-sight direction of movement from HH objects. The other is the diffuse 
redshifted component that appeared in Fig.\ref{outflowCM}. This complicated morphology makes it difficult to 
discuss the relation between IRS4 and HH49/50. To resolve this issue, one should enlarge CO observations
toward the southwest to know the larger structure of the outflow.

In Fig.\ref{pv7} and \ref{pv9}, the HCO$^+$ line toward IRS4 indicates a wide profile as well as a CO line. 
Infalling and outflowing gas seem to be the causes of this wide line \citep{Greg1997, Masunaga2000}, and 
perhaps, both the effects are blended. In both position-velocity diagrams, the blueshifted component of 
the HCO$^+$ line is broader than redshifted component. This is identical to the tendency of CO outflow, 
which indicates the wide HCO$^+$ line is mainly caused by the outflow. As in the HCO$^+$ channel map
(Fig.\ref{HCOchmap}), the blueshifted component extends to the west of IRS4, whereas redshifted component 
is elongated to the north. This aspect again corresponds with the morphology of the CO outflow 
(Fig.\ref{outflowCM}).  

Another method to use to determine what makes the wide line profile is to compare the velocity of 
infalling and outflowing gas. We calculated the free fall velocity of infalling gas, $v_{\rm ff} = 0.31$ -- 
0.56 km s$^{-1}$, assuming the mass range of 0.10 -- 0.32 $M_{\sun }$ derived by \citet{Persi2001}, and 
a diameter of the envelope of $1.7 \times 10^{-2}$ pc. These $v_{\rm ff}$ are 
smaller than the line width, which indicates the outflow has a larger contribution to the wide line 
profiles. 

Toward IRS6, we derived some physical parameters of the outflow in the same 
procedure as for IRS4 and show the result in Table. \ref{outflowtables}. In the case of IRS6, we could not 
correct the effect of the inclination of the outflow axis because of a lack of information. Besides, there have 
been no CO($J$=1-0) observations toward IRS6 in previous years, so we could not derive $\tau_{32}$. Here we use
the same $\tau_{32}$ as toward IRS4. Although there are ambiguities of inclination and optical depth, the 
outflow from IRS6 is more than 1 order of magnitude smaller than that
from IRS4. This is consistent with our estimation of its evolved status with a small envelope. This
interpretation is not consistent with the young age of the outflow; however, when the outflow is nearly in 
pole-on configuration, the dynamical timescale of the outflow will be calculated to be smaller than the
true value. We need to know the inclination angle of this source to correct this mismatch.

\subsection{Interaction between Cha-MMS1 and the Outflow from IRS4}
\begin{figure}
\epsscale{.80}
\plotone{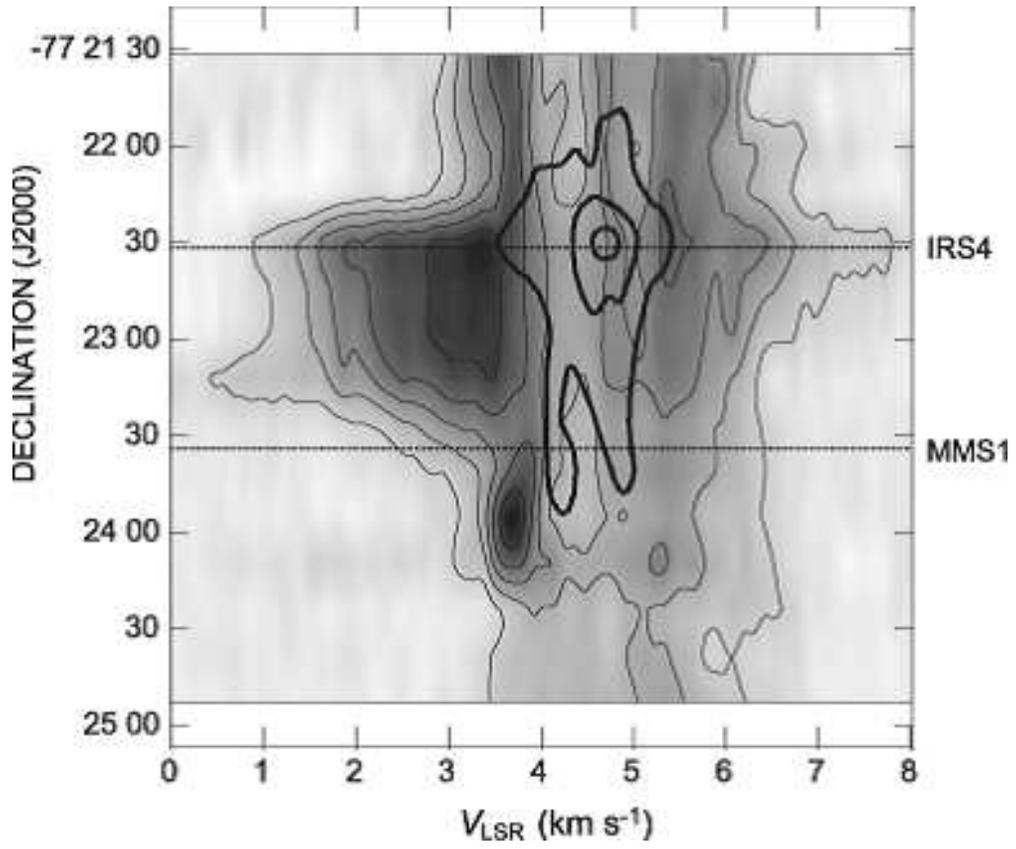}
\caption{Same as Fig. \ref{pv7}, but for line (b) in Fig. \ref{HCOmap}.
\label{pv9}}
\end{figure}

Figure \ref{outflowIT} shows the blueshifted component of the outflow from the IRS4 has negative spatial correlation with
the continuum emission from Cha-MMS1. The outflow seems to change its direction at the edge of the MMS1. In 
Fig.\ref{outflowCM}, the outflow component with the highest velocity is distributed at the midpoint between 
IRS4 and MMS1. In addition, the velocity of the outflow suddenly declines in front of MMS1 (Fig.\ref{pv9}), i.e., the 
flow seems to be decelerated by MMS1.  Do these characteristics indicate that MMS1 interacts with the 
outflow from IRS4? 

There were no emissions detected from shock tracer molecules in submillimeter wavelength in our observations.
However, considering IRS4's low luminosity \citep[$L_{\rm bol}=1.0 L_{\odot }$;][]{Lehtinen2001}, weak 
outflow properties, and high upper energy levels of $> 70$ K, it is conceivable that the outflow-core 
interaction could not increase the temperature enough to excite these molecules. We could not rule out that 
a weak interaction does occur. \citet{Kontinen2000} pointed out that outflows and radiation from the neighboring
young stars have prospects for influencing the chemical processes in MMS1.
Here we considered the impact of the interaction from following two view-points. 

First we investigate momentum. Here we compare the momentum brought by the outflow with gas momentum in MMS1.
When an outflow with density $\rho$ and velocity $v$ is colliding with a surface of radius $r$ for a time 
duration $t$, we can obtain the mass $\dot{M}_{\rm col}$ colliding in a unit of time and the momentum
$P_{\rm outflow}$ as,
\begin{eqnarray}
P_{\rm outflow} = \dot{M}_{\rm col} vt= \pi r^2v^2\rho t.
	 \label{outflowMom}
\end{eqnarray}
Now we estimate the density of the outflow. For simplicity, we assume the outflow has a cone shape with 
a base diameter of 40$''$ (= 3.1 $\times 10^{-2}$ pc) and a height of 60$''$ (= 4.7 $\times
10^{-2}$ pc) judging from Fig.\ref{outflowIT}. Calculating the density $\rho $ of the outflow from this 
assumption with the derived mass, we obtain $\rho =6.5 \times 10^{-20}$ g cm$^{-3}$. Assigning this 
density, radius $r = 1.6 \times 10^{-2}$ pc and $t$ which equals the dynamical timescale of the flow 
($1.0 \times 10^4$ yr) to equation (\ref{outflowMom}), we derived a momentum
of 0.88 $M_{\sun }$ km s$^{-1}$. Meanwhile, the momentum of gas in MMS1 is calculated to be
$P_{\rm MMS1} = 0.15$ $M_{\sun }$ km s$^{-1}$ from the mass of 0.45 $M_{\sun }$ and the velocity of 0.34 km
s$^{-1}$ derived from the H$^{13}$CO$^+$ line profile, which is a combination of turbulence and thermal width,
removing the contribution of the spectral resolution. As a result, $P_{\rm outflow}$ is much larger than
$P_{\rm MMS1}$, which means the motion of the
gas in MMS1 is easily affected when the outflow collide with the core.

Although the MMS1 contours in \citet{Reipurth1996} have an asymmetry, it is difficult to confirm the collisional
impact on the geometry of MMS1 with their map. But in our HCO$^+$ position-velocity diagram (Fig.\ref{pv9}), we 
could find the sign of the kinematic motion of the outer gas of MMS1 caused by the collision. The dip in the 
northern side of MMS1 is slightly ($\sim$ 0.3 km s$^{-1}$) blueshifted, which results in the velocity gradient
on the map. This gradient is consistent with the IRS4 outflow direction. Precise calculations of the momentum 
for this gradient is difficult with our data set, but in a rough upper limit estimation, the momentum for making 
this dip velocity gradient,
$P_{\rm grad}$, should be less than the product of the velocity difference and the whole mass of MMS1, 
$P_{\rm grad} < 0.14 M_{\odot }$ km s$^{-1}$. The momentum provided by the outflow is larger than 
$P_{\rm grad}$; thus, this outer gas motion could be excited by the interactions with the outflow.

The other point of discussion is the time scale before induced star formation is triggered in MMS1. One example of 
outflow-triggered
star formation is L1551 NE in Taurus \citep{Yokogawa2003}. In the case of MMS1, before star formation 
is triggered, it is necessary that a rise in pressure outside of MMS1 by colliding with the 
outflow reaches the center of the core. The velocity of this propagation equals the sound speed, or the shock 
speed if a shock has been excited. The sound speed under 10 K is 0.19 km s$^{-1}$. With this velocity, the 
signal of the collision arrives at the center in $7.2 \times 10^4$ yr. This is about an order of magnitude 
longer than the dynamical timescale of the outflow. Under this condition, triggered star formation would 
not have occurred yet. Next, we also consider the case in which a shock exists. In order to derive the
shock speed, we need to know the density structure around MMS1, because the shock speed depends on the 
difference between inner and outer density of MMS1. According to \citet{Reipurth1996}, the difference between 
the 1.3 mm continuum intensity at the center and outside of the MMS1 is about a factor of 2. Under optically 
thin conditiona, the difference of the intensity is identical to that of density, and the shock speed equals 
half of the outflow speed, 5.3 km s$^{-1}$. With this speed, the shock would arrive at the center in $2.8 \times 
10^3$ yr from the point of the collision. Considering the time to reach the boundary of MMS1 from its 
departure from IRS4, the time for the outflow to arrive at the center of MMS1 is $8.5 \times 10^3$ yr, 
about the same as the dynamical timescale of the outflow. If
the estimate of density structure is correct, compression has just reached to the center of MMS1 and 
collapse would just be triggered. Considering that we could not detect any infall and outflow signature 
around the source, it is possible that MMS1 is in a very young stage of its evolution.

\section{SUMMARY}

We observed an YSO condensation in the Cederblad 110 region of the Chamaeleon I dark cloud by CO($J$=3-2), 
HCO$^+$($J$=4-3), H$^{13}$CO$^+$($J$=4-3), CH$_3$OH($J$=$7_K$-$6_K$), SO($J$=$8_9$-$7_8$), and SiO($J$=8-7)
submillimeter lines with the ASTE 10m telescope. Results are as follows.

1) We found a dense HCO$^+$ clump. This clump is an aggregate made up of three components associated
with IRS4, IRS11, and Cha-MMS1. Although these sub-components do not appear in the integrated intensity map,
we can distinguish them in the channel map and the position-velocity diagrams.

2) We detected CO ($J=3-2$) outflow components around IRS4 and IRS6, but not 
around Cha-MMS1, IRS11, or NIR89. Our observations with high angular resolution 
show the outflows previously known in this region are launched by IRS4, not by MMS1. This result
suggests MMS1 is not a protostar but still in the prestellar phase. 

3) Each HCO$^+$($J$=4-3) line profile toward IRS4, IRS11, and MMS1 all have "red" asymmetry 
instead of blue asymmetry, which is a marker of infalling gas. However, our result does not necessarily mean
there is no infall in these young sources, because the outflows from IRS4 and IRS6 are direct evidences of 
gas infall.  

4) The blue-shifted outflow from IRS4 seems to collide against MMS1. The outflow
veers and the velocity declines just in front of MMS1. 
Though emission lines from shock tracer molecules such as CH$_3$OH, SO, and SiO are not detected, it is
plausible because the outflow from IRS4 is rather weak for excitation of these molecules.

5) Momentum calculations show the outflow from IRS4 is strong enough to affect gas motion in MMS1, 
although the outflow is so young that the star formation process has not been triggered yet, or has just
been triggered. This is consistent with the very young status of MMS1 and the lack of any outflows, jets, and
masers.

\acknowledgments
We are grateful to all ASTE project members, particularly Takeshi Kamazaki and Kazuyoshi Sunada, 
for supporting the dual-frequency observations. Observations with ASTE were in part carried out remotely
from Japan by using NTT's GEMnet2 and its partner R\&E (Research and Education) networks, which are based on
AccessNova collaboration of University of Chile, NTT Laboratories, and the National Astronomical Observatory
of Japan (NAOJ). ASTE is a subproject of the
Nobeyama Radio Observatory which is a branch of NAOJ, 
operated by the Ministry of Education, Culture, Sports, Science, and Technology (MEXT), Japan.
This project was financially supported by MEXT Grant-in-Aid for Scientific Research
on Priority Areas 15071202. Data analyses were carried out on the general common-use computer system 
at the Astronomy Data Center of NAOJ.

\clearpage

\begin{deluxetable}{lllllll}
\tablecolumns{7}
\tablewidth{0pc}
\tablecaption{YSOs in Cederblad 110 Region}
\tablehead{
	\colhead{Name} & \colhead{R.A. (J2000)} & \colhead{Dec. (J2000)} & \colhead{IR class}
	 & \colhead{$T_{\rm bol}$} & \colhead{$L_{\rm bol}$} & \colhead{References}
} 
\startdata
	Cha-MMS1    & 11 06 31.7 & -77 23 32 & 0 & 20 & 0.46 & 1\\
	 &  &  & 0 & 36 & 0.38 &2 \\
	Ced110 IRS4 & 11 06 47.1 & -77 22 34 & I & 72 & 1.0 & 1\\
	 &  &  & 0/I & 59 & 1.0\ &2 \\
	Ced110 IRS6a & 11 07 9.8 & -77 23 05 & I & 260 & 0.8 & 1\\
	Ced110 IRS11 & 11 06 58.8 & -77 22 50 & I & \nodata & 0.17 & 3\\
	NIR89 & 11 06 53.8 & -77 24 00 & I & \nodata & \nodata& \\
	ISO-Cha I 97  & 11 07 17.0 & -77 23 08 & II & \nodata & \nodata& \\
	ISO-Cha I 101 & 11 07 21.7 & -77 22 12 & II & \nodata & \nodata& \\
	Ced110 IRS2  & 11 06 16.8 & -77 21 55 & III & 2700 & 3.4 & 1 \\
\enddata
\tablerefs{(1) \citet{Lehtinen2001}; (2) \citet{Froebrich2005}; (3) \citet{Persi2001}}
\label{Ced110table}
\end{deluxetable}

\begin{deluxetable}{llc}
\tablecolumns{3}
\tablewidth{0pt}
\tablecaption{CO($J$=3-2) Observations on YSOs}
\tablehead{
	\colhead{} & \colhead{$T_{\rm MB}$} & \colhead{$\int T_{\rm MB} {\rm d}v$} \\
	\colhead{} & \colhead{(K)}  & \colhead{(K km s$^{-1}$)} }
\startdata 
	Cha MMS1    & 6.94 $\pm$ 0.21 & 14.4 $\pm$ 0.1 \\
	Ced110 IRS4 & 8.39 $\pm$ 0.43 & 34.7 $\pm$ 0.1 \\
	Ced110 IRS6 & 11.37 $\pm$ 0.20 & 24.3 $\pm$ 0.1 \\
	Ced110 IRS11 & 6.50 $\pm$ 0.47 &15.5 $\pm$ 0.1\\
	NIR89        & 7.86 $\pm$ 0.41 & 17.6 $\pm$ 0.1 \\
	ISO-ChaI 97  & 8.22 $\pm$ 0.21 & 16.0 $\pm$ 0.1 \\
	ISO-ChaI 101 & 6.94 $\pm$ 0.27 & 14.1 $\pm$ 0.1 \\
	Ced110 IRS2  & 7.80 $\pm$ 0.38 & 14.9 $\pm$ 0.1 \\
\enddata
\label{COtable}
\end{deluxetable}

\begin{deluxetable}{llccc}
\tablecolumns{5}
\tablewidth{0pt}
\tablecaption{HCO$^+$($J$=4-3) Observations on YSOs}
\tablehead{
	\colhead{} & \colhead{$T_{\rm MB}$} & \colhead{$V_{\rm LSR}$\tablenotemark{a}} &\colhead{$\Delta V$\tablenotemark{b}}
	 & \colhead{$\int T_{\rm MB} {\rm d}v$} \\
	\colhead{} & \colhead{(K)} & \colhead{(km s$^{-1}$)} & \colhead{(km s$^{-1}$)} & \colhead{(K km s$^{-1}$)} }
\startdata 
	Cha MMS1    & 1.64 $\pm$ 0.09 & 4.49 $\pm$ 0.12 & 0.68 $\pm$ 0.14 & 1.11 $\pm$ 0.01 \\
	Ced110 IRS4 & 3.00 $\pm$ 0.09 & 4.46 $\pm$ 0.13 & 1.55 $\pm$ 0.15 & 4.58 $\pm$ 0.01 \\
	Ced110 IRS6 & 0.65 $\pm$ 0.09 & 4.62 $\pm$ 0.18 & 1.53 $\pm$ 0.28 & 0.73 $\pm$ 0.01 \\
	Ced110 IRS11 & 1.47 $\pm$ 0.08 & 4.67 $\pm$ 0.13 & 0.90 $\pm$ 0.15 & 1.35 $\pm$ 0.01 \\
	NIR89        & $<$ 0.16 & \nodata & \nodata & $<$ 0.35 \\
	ISO-ChaI 97  & $<$ 0.17 & \nodata & \nodata & $<$ 0.70 \\
	ISO-ChaI 101 & $<$ 0.36 & \nodata & \nodata & $<$ 0.54 \\
	Ced110 IRS2  & $<$ 0.29 & \nodata & \nodata & $<$ 0.30 \\
\enddata
\tablenotetext{a}{$V_{\rm LSR}$ at the peak of the best-fit Gaussian profile.}
\tablenotetext{b}{FWHM of the best-fit Gaussian profile.}
\label{HCOtable}
\end{deluxetable}

\begin{deluxetable}{lcccc}
\tablecolumns{5}
\tablewidth{0pt}
\tablecaption{H$^{13}$CO$^+$($J$=4-3) Observations on YSOs}
\tablehead{
	\colhead{} & \colhead{$T_{\rm MB}$} & \colhead{$V_{\rm LSR}$\tablenotemark{a}} & \colhead{$\Delta V$\tablenotemark{b}}
	 & \colhead{$\int T_{\rm MB} {\rm d}v$} \\
	\colhead{} & \colhead{(K)} & \colhead{(km s$^{-1}$)} & \colhead{(km s$^{-1}$)} & \colhead{(K km s$^{-1}$)}}
\startdata 
	Cha MMS1 & 0.26 $\pm$ 0.03 & 4.63 $\pm$ 0.13 & 0.36 $\pm$ 0.16 & 0.11 $\pm$ 0.01 \\
	Ced110 IRS4 & 0.13 $\pm$ 0.03 & 4.53 $\pm$ 0.18 & 1.72 $\pm$ 0.29 & 0.18 $\pm$ 0.01 \\
\enddata
\tablenotetext{a}{$V_{\rm LSR}$ at the peak of the best-fit Gaussian profile.}
\tablenotetext{b}{FWHM of the best-fit Gaussian profile.}
\label{H13COtable}
\end{deluxetable}

\begin{deluxetable}{lcccc}
\tablecolumns{4}
\tablewidth{0pt}
\tablecaption{H$_2$ Column Density and HCO$^+$ Abundance toward IRS4 and MMS1}
\tablehead{
	\colhead{} & \colhead{$S^{\rm beam}_{\rm 1.3 mm}$} & \colhead{$T_{\rm D}$} & \colhead{$N$(H$_2$)} & \colhead{$X$(HCO$^+$)} \\
	\colhead{} & \colhead{(mJy)} & \colhead{(K)} & \colhead{(cm$^{-2}$)} & \colhead{ }}
\startdata 
	Cha MMS1 & 370\tablenotemark{(1)} & 20\tablenotemark{(3)}    & $2.2 \times 10^{23}$ & $2.6 \times 10^{-10}$   \\
	Ced110 IRS4 & 101\tablenotemark{(2)} & 21\tablenotemark{(3)} & $2.8 \times 10^{22}$ & $3.4 \times 10^{-9}$  \\
\enddata
\tablerefs{(1) \citet{Reipurth1996}; (2) \citet{Henning1993}; (3) \citet{Lehtinen2001}}
\label{NH2table}
\end{deluxetable}

\begin{deluxetable}{cccccccc}
\rotate
\tablecolumns{8}
\tablewidth{0pt}
\tablecaption{Parameters for Outflows from Ced 110 IRS4 and IRS6}
\tablehead{
	\colhead{} & \colhead{$R_{\rm outflow}$} & \colhead{$M$} & \colhead{$t_{\rm d}$}
	 & \colhead{$\dot{M}$} & \colhead{$P$} & \colhead{$E_k$} & \colhead{$F$}\\
	\colhead{} & \colhead{(pc)} & \colhead{($M_{\sun }$)} & \colhead{(year)}
	 & \colhead{($M_{\sun }$yr$^{-1}$)} & \colhead{($M_{\sun }$km s$^{-1}$)}
	 & \colhead{($M_{\sun }$(km s$^{-1}$)$^2$)} & \colhead{($M_{\sun }$ km s$^{-1}$ yr$^{-1}$)} }
	\startdata
	\sidehead{IRS4} 
	Blue & 6.5$\times 10^{-2}$ & 1.1$\times 10^{-2}$ & 1.0$\times 10^4$ & 1.1$\times 10^{-6}$ & 7.1$\times 10^{-2}$ & 2.4$\times 10^{-1}$ & 7.1 $\times 10^{-6}$ \\
	Red & 2.5$\times 10^{-2}$ & 9.8$\times 10^{-4}$ & 2.9$\times 10^3$ & 3.4$\times 10^{-7}$ & 7.7$\times 10^{-3}$ & 3.1$\times 10^{-2}$ & 2.7 $\times 10^{-6}$ \\
	\sidehead{IRS6} 
	Blue & 1.6$\times 10^{-2}$ & $3.3\times 10^{-4}$ & 7.8$\times 10^3$ & 4.3$\times 10^{-8}$ & 6.5$\times 10^{-4}$ & 6.7$\times 10^{-4}$ & 8.4 $\times 10^{-8}$ \\
	Red  & 3.1$\times 10^{-2}$ & $5.7\times 10^{-4}$ & 1.3$\times 10^4$ & 4.3$\times 10^{-8}$ & 1.3$\times 10^{-3}$ & 1.5$\times 10^{-3}$ & 9.9 $\times 10^{-8}$ \\
\enddata
\label{outflowtables}
\end{deluxetable}

\clearpage

\end{document}